\documentclass[conference]{IEEEtran}
\usepackage{verbatim}
\usepackage{longtable}
\usepackage{xcolor}
\usepackage{amsmath}
\usepackage[ruled,vlined,linesnumbered]{algorithm2e}
\usepackage{stfloats}
\usepackage{etoolbox}
\usepackage{graphicx}
\BeforeBeginEnvironment{figure}{\vskip-2ex}
\AfterEndEnvironment{figure}{\vskip-1ex}
\usepackage{float}
\usepackage{algorithmicx}
\usepackage{fancyhdr}
\usepackage{cite}

\usepackage{amsmath}
\usepackage{longtable}

\usepackage{dirtytalk}
\ifCLASSOPTIONcompsoc
\usepackage[caption=false,font=normalsize,labelfont=sf,textfont=sf]{subfig}
\else
\usepackage[caption=false,font=footnotesize]{subfig}
\fi
\begin{document}
\pagestyle{empty}
\title{\textbf{\texttt{FASTEN:}} Towards a \textbf{\texttt{FA}}ult-tolerant and \textbf{\texttt{ST}}orage \textbf{\texttt{E}}fficie\textbf{\texttt{N}}t Cloud: Balancing Between Replication and
Deduplication}

\author{\IEEEauthorblockN{Sabbir Ahmed$^{1}$, Md Nahiduzzaman$^{2}$,  Tariqul Islam$^{3}$, Faisal Haque Bappy$^{4}$, \\ Tarannum Shaila Zaman$^{5}$, and Raiful Hasan$^{6}$}
\IEEEauthorblockA{
$^{1, 2}$ Institute of Information Technology, Jahangirnagar University, Dhaka, Bangladesh \\
$ ^{3, 4}$ School of Information Studies (iSchool), Syracuse University, Syracuse, NY, USA\\
$ ^{5}$ Computer and Information Science, SUNY Polytechnic Institute, NY, USA\\
$ ^{6}$ Computer Science, Kent State University, Kent, OH, USA\\
Email: \{sabbir.iit.ju@, nahidsamrat2@\}gmail.com; \{mtislam@, fbappy@\}syr.edu and \\ \{zamant@sunypoly, rhasan7@kent\}.edu}}

\maketitle

\thispagestyle{fancy}
\lhead{This work has been accepted at the 2024 IEEE Consumer Communications \& Networking Conference (CCNC 2024)}
\cfoot{}
\begin{abstract}
With the surge in cloud storage adoption, enterprises face challenges managing data duplication and exponential data growth. Deduplication mitigates redundancy, yet maintaining redundancy ensures high availability, incurring storage costs. Balancing these aspects is a significant research concern. We propose \texttt{FASTEN}, a distributed cloud storage scheme ensuring efficiency, security, and high availability. \texttt{FASTEN} achieves fault tolerance by dispersing data subsets optimally across servers and maintains redundancy for high availability. Experimental results show \texttt{FASTEN}'s effectiveness in fault tolerance, cost reduction, batch auditing, and file and block-level deduplication. It outperforms existing systems with low time complexity, strong fault tolerance, and commendable deduplication performance.
\end{abstract}


\section{Introduction}
In the era of big data, cloud computing revolutionizes data sharing among owners and authorized users, reshaping enterprise strategies in hardware, software design, 
and procurement. Users' demand for cloud services increases as data volumes grow, while providers navigate maintaining system availability amidst this data influx. Cloud 
service providers (CSPs) seek methods like data deduplication \cite{yuan2018dedupdum} to trim data volume, reducing storage costs and bandwidth. However, deduplication 
may compromise availability and reliability, essential for cloud services. Replication stands out in cloud computing to ensure data accessibility, safety, and security 
across servers. Yet, excessive replicas inflate storage costs. Our aim is to strike a balance between deduplication and replication, achieving an efficient, highly 
reliable storage system without excessive expense.

Cloud storage systems employ data replication to distribute data intelligently among multiple cloud providers, enhancing availability. Various strategies \cite{abu2010racs}, \cite{bowers2009hail}, \cite{hu2012nccloud}, \cite{steans2015duracloud}, \cite{bessani2013depsky} propose duplicating user data across providers. HyRD \cite{mao2015improving} integrates erasure coding with a replication strategy for efficient, highly available storage. Studies by Microsoft \cite{meyer2012study}, EMC \cite{xia2016comprehensive}, and IDC \cite{dubois2011key} reveal redundancy levels in production and backup storage, advocating data deduplication for enhanced storage efficiency and cost reduction. Douceur \cite{douceur2002reclaiming} introduced convergent encryption for secure data deduplication, with various methods \cite{jayapandian2018secure}, \cite{yang2020achieving}, \cite{singh2018secure}, \cite{yuan2019secure}, \cite{li2019csed} based on this concept in deployment or planning stages.

Prior research (RACS \cite{abu2010racs}, HAIL \cite{bowers2009hail}, DuraCloud \cite{steans2015duracloud}, NCCloud \cite{hu2012nccloud}, DepSky \cite{bessani2013depsky}, and HyRD \cite{mao2015improving}) favors replication-based schemes for enhanced performance in availability and reliability, while deduplication-based schemes \cite{jayapandian2018secure}, \cite{yang2020achieving} are cost-effective due to reduced redundancies. Enterprises can benefit from client-side data deduplication before cloud outsourcing to save costs. Our proposed algorithms aim to ensure fault tolerance, consistency-checked data deduplication, improved availability, and security. The following are the major contributions of the paper.

\begin{itemize}
    \item We introduce \texttt{FASTEN}, a novel cloud storage data dispersal scheme that balances ``deduplication" and ``replication".
    \item Our contributions include the \texttt{fault-tolerant subset} and \texttt{server rating} algorithms. The former organizes data blocks for maximum fault tolerance, while the latter selects optimal servers based on user-defined redundancy.
    \item Our prototype of \texttt{FASTEN} measures performance metrics (read, write, update, auditing, fault tolerance) across varying file/block sizes and redundancy factors. We compare it against state-of-the-art deduplication, fault tolerance, and batch auditing schemes.
    \item We designed \texttt{write} and \texttt{update} algorithms in \texttt{FASTEN} to manage file and block-level deduplication while ensuring data security and privacy. 
    \item We incorporated \texttt{batch auditing} in our scheme using two data structures, i) our custom \texttt{HashMap} and ii) \texttt{MHT} technique. We then compared batch auditing performance between them. 

 
    
\end{itemize}

The rest of the paper is organized as follows: in Section \ref{Preliminaries}, we review some preliminaries and cryptographic primitives along with a few well-known security algorithms and protocols. In Section \ref{model}, we present our proposed scheme in detail, followed by experimental evaluations in Section \ref{experimental}. We compare and contrast existing work with our work in Section \ref{related}. Finally, we conclude the paper in Section \ref{conclusion}.

\section{Preliminaries} \label{Preliminaries}


\textbf{Convergent Encryption and Deduplication.} Convergent Encryption (CE) generates identical ciphertext for duplicate files using a convergent key derived from the data's hash. If a user attempts to upload duplicate data, the server discards it and issues an ownership pointer. CE ensures data confidentiality by encrypting message blocks using a convergent key.

\textbf{Merkle Hash Tree.} A Merkle hash tree is a hash-based data structure for efficient digital data authentication. Internal nodes store concatenated hash values of their left and right children. By dividing a file into blocks, coupling them, and iteratively hashing pairs with a collision-resistant function, a Merkle tree is created. This process continues until only one hash value, the root, remains.




\textbf{Hashmap.} Data structures with indexes are hash maps. When creating an index with a key into an array of buckets or slots, a hash map uses a hash function. The bucket with the relevant index is linked to its value. The key is distinct and unchangeable. Hash maps include the following functions: i) \texttt{SetValue(key, value)}: A key-value pair is inserted into the hash map. This function updates the value if it is already there in the hash map; ii) \texttt{GetValue(key)}: If there is no mapping for the given key in this map, this method returns ``No record found" or returns the value to which the given key is mapped and iii) \texttt{SetValue(key)}: Deletes the mapping for a certain key if it is present in the hash map.

\begin{table}[ht]
\centering
\caption{Notations for Protocol Flow}
\label{table:notation}
\resizebox{\columnwidth}{!}{%
\begin{tabular}{|c|c|c|c|}
\hline
\textbf{Notation} &
  \textbf{Description} &
  \textbf{Notation} &
  \textbf{Description} \\ \hline
$U_{id}$ &
  User ID &
  $MHT$ &
  Merkle Hash Tree \\ \hline
$A$ &
  Storage Address &
  $M_{r}$ &
  Merkle Hash Root \\ \hline
$S$ &
  Data Server &
  $K_{CE}$ &
  Convergent Key \\ \hline
$S_{id}$ &
  Data Server Id &
  $C_i$ &
  Ciphertext \\ \hline
$A_v$ &
  Availability &
  $nS_{m}$ &
  \begin{tabular}[c]{@{}c@{}}No. of Maximum \\ Allocated Servers\end{tabular} \\ \hline
$IS$ &
  Index Server &
  $S_{op}$ &
  Optimum Data Servers \\ \hline
$F_j$ &
  Files &
  $divs_i$ &
  Divisors \\ \hline
$F_{id}$ &
  File id &
  $Av_{s}$ &
  Available Space \\ \hline
$H_k$ &
  Tags &
  $S_l$ &
  Server Load \\ \hline
$B$ &
  Data Block &
  $R_f$ &
  Redundancy Factor \\ \hline
$B_{sz}$ &
  Size of Data Block &
  $B_{ss}$ &
  Data Block Subset \\ \hline
$B_{con}$ &
  \begin{tabular}[c]{@{}c@{}}Concatenated \\ Data Block\end{tabular} &
  $H_{b}$ &
  Hash of Data Blocks \\ \hline
$H_{con}$ &
  Concatenated Tag &
  $H_{ss}$ &
  Hash Subset \\ \hline
$B_{ss}$ &
  \begin{tabular}[c]{@{}c@{}}Subset of Data \\ Block\end{tabular} &
  $dif_{h}$ &
  \begin{tabular}[c]{@{}c@{}}Difference of Current \\ and Previous Hash\end{tabular} \\ \hline
$H_{ss}$ &
  Subset of Tags &
  $DP$ &
  Memorization Table \\ \hline
$HM$ &
  HashMap &
  $DD$ &
  \begin{tabular}[c]{@{}c@{}}Deduplicated \\ Redundancy Score Table\end{tabular} \\ \hline
\end{tabular}%
}
\end{table}

\section{System Model and Design Principles} \label{model}

\begin{figure}[b]
\centering
\includegraphics[width=3.1in,height=2.2in]{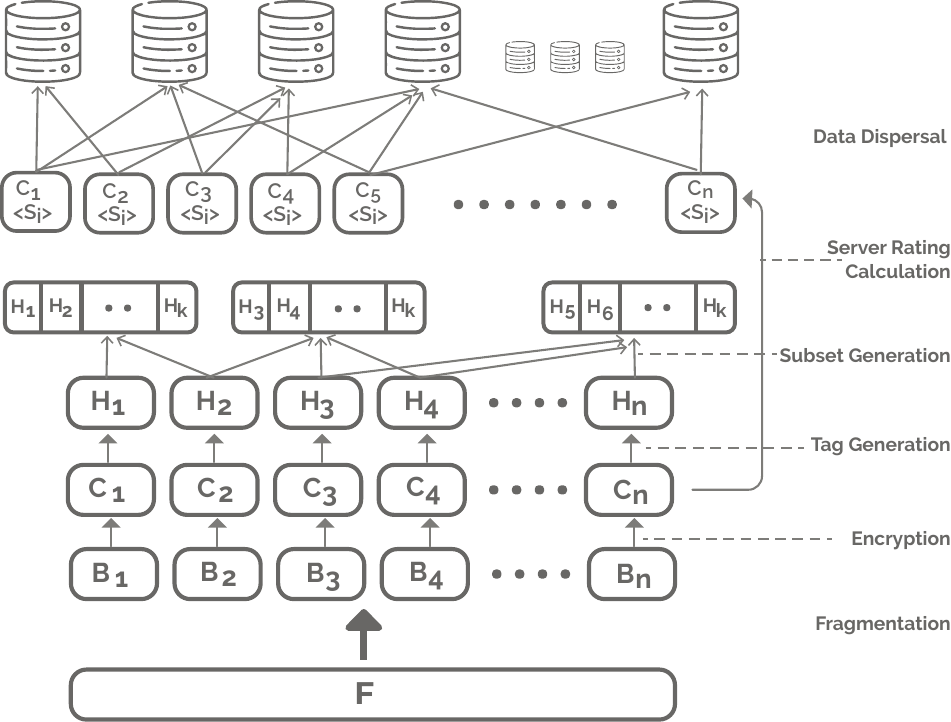}
\caption{System Architecture}
\label{SysMod}
\end{figure}

In this section, first, we define several data structures and functionality of the proposed \texttt{FASTEN} system, which will be later utilized by various user-level operations to perform \texttt{read}, \texttt{write}, \texttt{update}, and \texttt{delete} operations. A high-level overview of the whole process is shown in Fig. \ref{SysMod} and the notations used in our algorithms are listed in Table \ref{table:notation}.

\subsection{Index Server Structure and Initialization}
The Index server additionally keeps track of a collection of information regarding users, data servers, and previously uploaded data. The structures of the stored data are as follows.

\textbf{Cloud Users.} \textit{Users (U)} is an unordered map structure with unique user IDs ($U_{id}$) containing multiple files ($F$). Each file has ordered hash lists ($H_k$). Any number of $U_{id}$, $F_{id}$, or $H_k$ can be inserted, replaced, or deleted in constant time complexity. \\


\textbf{Data Server.} The data server $S$ is an unordered structure tracking server availability and storage addresses ($A$). Multiple servers ($S_{i}$) are identified by unique keys ($S_{id}$). In each $S_{id}$, a boolean list is created (i.e., \texttt{TRUE: not available; FALSE: space available}) with the corresponding storage location $A$ as key. 


\textbf{HashMap.} Our \texttt{HashMap(HM)} is a simple but effective unordered map structure with a hash value ($H_k$) as the key, where each $H_k$ is an authentication tag and also a general purpose tag for data block ($B_i$). For each $H_k$, several pairs of ($S_{{id}_j}$, $A_j$) can be assigned to describe the exact memory location and server identification for data storing and retrieval. 

\textbf{MHT Construct.} To verify data integrity, we construct a Merkel Hash Tree (MHT) using our structure. Initially, a Merkle root $M_{r}$ is created from any file $F$ using the Merkle tree technique. After uploading all file blocks to cloud servers, the user (or an auditor) can query the server to prove the verifiability of specific data blocks. The server responds with its answer to the challenge query. The user recalculates the Merkle root using the server's response and checks whether it matches its stored root to verify data integrity.


\subsection{Data Processing}
Our system utilizes the following methods for data processing: i) convergent encryption generates keys by hashing each file; ii) AES-256 symmetric cryptography encrypts each file; iii) files are fragmented into data blocks based on a specified block size, and iv) authentication tags are generated for each block by hashing with the SHA-256 algorithm.


\subsection{Applying Redundancy}
\subsubsection{Optimum Servers and Subsets}
Since blocks can be sent to any number of available data servers $S$, the exact number of optimum servers needs to be calculated from the maximum servers allocated to any particular user and the user-defined redundancy factor. Given a number of allocated servers $nS_{m}$, for any particular user, with the number of total data blocks $n_B$ for a specific file and redundancy factor $R_f$, here we find out the optimum number of servers $nS_{op}$ that can be used to store these data blocks.

\subsubsection{Maximum Fault Tolerance Subset (FT Subset)}
Data blocks $B_i$ may have several combinations of subsets when we consider the redundancy (replication) factor ($R_f$). Hence, in this section, our goal is to generate maximum fault-tolerant subsets of blocks and tags while considering the user-defined redundancy factor ($R_f$). 

\textbf{Problem Definition:} $R$ copies of $n_B$ data blocks need to be stored in $nS_{op}$ servers. What could be the most efficient and fault-tolerant way to do that?

\textbf{Assumption :} If we can disperse data blocks evenly among servers, that could reduce the points of failure. Therefore, the length (i.e., size) of data blocks sent to each server must be the same to achieve an even distribution. 

\IncMargin{1em}
\setlength{\textfloatsep}{0pt}
\begin{algorithm}
\SetKwInOut{Input}{input}\SetKwInOut{Output}{output}
\Input{$\{B_i\}$, $\{H_k\}$, $nS_{op}$, $R$ \Comment{blocks, tags, \# optimum servers and redundancy factor}}

\Output{$\{H_{ss}\}$,$\{B_{ss}\}$ \Comment{subsets of tags and blocks}}
\BlankLine

{$size$ $\leftarrow$ \texttt{length($\{H_k\}$)};}
\Comment{length of a tag-subset}

$DBCon[\ ]$ $\leftarrow$ \texttt{init()}\;

$HCon[\ ]$ $\leftarrow$ \texttt{init()}\;
$ssz$ $\leftarrow$ \texttt{($size$ $\times$ $R$)$/$$nS_{op}$}
\Comment{subset size}\\

\If{\texttt{isNotInt}$(ssz)$}
{
 \texttt{return $``Error"$}\;
}

\For{$i \gets 1$ to $R$}{
$B_{con}$.\texttt{add} (\texttt{$\{B_i\}$}); \Comment{block replication} \\
$H_{con}$.\texttt{add} (\texttt{$\{H_k\}$}); \Comment{tag replication} \\
} 
\For{$j\gets0$ \KwTo ($size$ $\times$ $R$)}
{
$\{B_{ss}\}$ $\leftarrow$ \texttt{split}(\texttt{$B_{con},j,ssz)$}; \Comment{block subset}\\ 
$\{H_{ss}\}$ $\leftarrow$  \texttt{split}(\texttt{$H_{con},j,ssz)$}; \Comment{tag subset}\\
$j+=ssz;$  \Comment{advance to next subset}\\
}

 \texttt{return $\{H_{ss}\}$, $\{B_{ss}\}$}\;

\caption{\textbf{
\texttt{maxFTSubset($\{B_i$\},$\{H_k\}$,$nS_{op}$,$R$)}}}
\label{algo_3}
\end{algorithm}
\DecMargin{1em}

\textbf{Postulate 2}: For any arbitrary value of  $nS_{op}$, $n_B$, and $R_f$, the possibility of even block distribution can be checked (Algorithm-\ref{algo_3} lines 4-6). Otherwise,  the list of servers $S$  can be chosen from the set of the proper divisors of $(n_B*R_f)$. Since the redundancy factor is $R_f$, each of the data blocks and its associated hash tags (i.e., authentication tags) must occur exactly $R_f$ times in a concatenated state (Algorithm-\ref{algo_3}, lines 8-10). Finally, the algorithm calculates the final subsets of blocks and tags in lines 12-15. Since there are $n_B$ data blocks and they are arranged sequentially, the maximum distance between the two exact copies is $n_B$, which is achieved in this solution. Here, two copies of the same data block are not placed in one subset. If they were placed in one subset, then by losing just one subset, that part of the data could be lost, which would reduce the fault tolerance.

\IncMargin{1em}
\setlength{\textfloatsep}{0pt}
\begin{algorithm}
\SetKwInOut{Input}{input}\SetKwInOut{Output}{output}
\Input{$\{H_{ss}\}$,$\{S_i\}$ \Comment{subset of tags, list of servers}}

\Output{$\{H_{ss}$, $S_{i}\}$ \Comment{optimum pair (tag subsets, servers) alignment}}
\BlankLine

$\{DP\},\{DD\}$ $\leftarrow$ \texttt{$0$};

\For{each $i$ $\in$ $\{H_{ss}\}$}{

    \For{each $j$ $\in$ $\{S_i\}$}{
    $\{DD_{i,j}\}$ $\leftarrow$ \texttt{$\{DD_{i,j}\}$ + findLoc($H_k$)};\\
    $\{Rc_{i,j}\}$ $\leftarrow$ \texttt{available($S_j$)};\\
    $\{S_l, Q_s, Dis\}_{i,j}$ $\leftarrow$ \texttt{getVal($S_j$)};\\
    
    }
}
$\{R_{i,j}\}$ $\leftarrow$ \texttt{weightedRating($\{DD_{i,j}\}$, $\{Rc_{i,j}\}$, $\{{S_l}_{i,j}\}$, $\{Qs_{i,j}\}$, $\{Dis_{i,j}\}$)};\\

\For{each $i$ $\in$ $\{H_{ss}\}$}{
    \For{each $j$ $\in$ $\{S_i\}$}{
    $\{DP_{i,j}\}$ $\leftarrow$ \texttt{$\{R_{i,j}\}$ + findMax(${DP}, i+1, j\pm1$)} \\
    }}

$\{H_{ss},S_i\}$ $\leftarrow$ \texttt{pathPrint($\{DP\}$)};\\
    
 \texttt{return $\{H_{ss}, S_{i}\}$}\;
   
\caption{\textbf{
\texttt{rating($\{H_{ss}\}$,$\{S_i\}$)}}}
\label{algo_rate}
\end{algorithm}
\DecMargin{1em}

\subsection{Server Rating Calculation} 

The subsets of data block $B_{ss}$ need to be distributed in $nS_{m}$ data servers where $nB_{ss} <= nS_{m}$; also, two subsets can not be assigned to the same data server $S_i$ for ensuring fault tolerance. For each subset of data blocks $B_{ss}$, our algorithm selects the best server $S_i$ that maximizes the overall rating score.

\textbf{Duplication Matching.}
For each data block $B_i$ in $B_{ss}$ subsets, the corresponding tag or hash ($H_k$) is calculated. Again, for any key $H_k$, Hashmap ($HM$), gives the server ID ($S_{id}$) and storage location ($A_j$) pairs to retrieve the data block ($B_i$) having tag $H_k$ (Algorithm-\ref{algo_rate}, lines 2-7). The unavailability of $H_k$ in $HM$ also can be queried in O(1) time. For each block-subset $B_{ss}$, duplicate count for each data server $S_i$ is calculated and stored in tabular form, where the $B_{ss}$’s $j^{th}$ row and $S_i$'s $i^{th}$ column represent the duplicate count for the corresponding $B_{ss}$ and $S_i$.

\textbf{Weighted Rating (WR).}
Weighted final rating (Algorithm-\ref{algo_rate}, line 9) can be computed by taking the weighted sum of all of the scoring criteria as equation (1): 

$WR$ $=$ $DD$*$a$ + $S_l$*$b$ + $Q_s$*$c$ + $Dis$*$d$ + $R_c$*$e$     (1)  \\ \vspace{-2mm}

where, $DD$ = deduplicated redundancy score, $S_l$ = server load, $Q_s$ = user-specified query size, $dis$ = distance, $R_c$ = remaining capacity of the server. Here ($a, b, c, d$) are numbers from series like $alpha*(1/2)^x$ or any arbitrary number. For the system, the constant that would be multiplied with each criterion is set as a decreasing geometric series, such that the first criterion carries the most weight in scoring. Since these criteria can vary depending on system requirements i.e., some systems, for instance, might require higher weight on server load than other factors. The total rating forms a 2D matrix where each of the $R_{ij}$   represents the rating for sending $i^{th}$ subset to $j^{th}$ data server where $0<i<=nB_{ss}$ and $0<j<=nS_i$. Thus, there are $nB_{ss}$ * $nS_i$ states from which we have to take the maximum pair only once for a block subset. This way we calculate all the ratings for the servers and subsets and memorize them on $DP$($B_{ss}$, $S_i$) table (Algorithm-\ref{algo_rate}, lines 8-10). To find out the optimal $S_i$ for each $B_{{ss}_j}$, we need to follow the path for the maximum value of $DP$ so that the path will eventually give unique ($B_{{ss}_j}$, $S_i$) pairs. This process of finding the maximum value and path is similar to the well-known ``Stable Marriage Dynamic Programming (DP)" problem \cite{gale1962college} where a stable matching or the maximum score between two sets of nodes is calculated. Each set has a preference for another set. Similarly in our case, $B_{ss}$ and $S_i$ are two different sets that have preference values (Weighted Rating) and the overall rating needs to be maximized. This is solved using Gale–Shapley algorithm\cite{gale1962college} with the complexity of O($n^2$).

\subsection{Data Dispersal}
In a shared multi-user environment, when a user wants to upload a file $F$, there can be two scenarios: i) the file is being uploaded for the first time, or ii) the file has already been uploaded by another user from the same group. 
\IncMargin{1em}
\setlength{\textfloatsep}{0pt}
\begin{algorithm}
\SetKwInOut{Input}{input}\SetKwInOut{Output}{output}
\Input{$F_{id}$, $B_{sz}$ \Comment{file and block size}}
\Output{$\{H_{k}$, $S_{i}\}$ \Comment{tags and data servers}}
\BlankLine
$\{B_{i}, H_{k}\}$ $\leftarrow$ 
\texttt{dataProcess($F$,$B_{sz}$)};\\

$\{S_{i}\}$, $nS_{op}$  $\leftarrow$ \texttt{getServerList($n_B$, $nS_m$)};\\

$\{H_{ss}$\}, \{$B_{ss}\}$ $\leftarrow$ \texttt{maxFTSubset($\{B_i\}$,$\{H_k\}$,$nS_{op}$,$R$)};\\

$\{H_{ss}, S_i\}$ $\leftarrow$ \texttt{rating($\{H_{ss}\}, \{S_i\}$)};\\

\For{each $\{B_i,H_k\}$}{
$\{isDup_i\}$ $\leftarrow$ \texttt{HashMap($\{H_{k}\}$)}\;
            
\If{$isDup_i$=\texttt{TRUE}}
{
 \texttt{return blockPointer($H_{k}$)}\;
}
{
  $A_j$ $\leftarrow$\texttt{available($S_i$)}\Comment{if any server has space available, return storage address or print ``error"}\;

  \texttt{upload($B_{i}$,$A_{j}$,$S_i$)}\;
  \texttt{dataServer[$S_i$][$A_{j}$]}$\leftarrow$\texttt{True}\;
  \texttt{updateHashMap()}\;
  \texttt{returnTag($F_{id}$)};
}}

\caption{\textbf{
\texttt{firstUpload($F_{id}$,$B_{sz}$)}}}
\label{algo_write}
\end{algorithm}
\DecMargin{1em}

\textbf{First Upload.} To write (upload to the cloud) any file, the user first takes an authentication key using the key generation function, then encrypts the data using ``convergent encryption" using the AES (Advanced Encryption Standard) algorithm. The encrypted data is sent to the index server (IS) with $U_{id}$, $F_{id}$ with some additional parameters like the block size $B_{sz}$, redundancy factor $R_f$, no. of maximum allocated servers $nS_{m}$, etc. The IS authenticates $U_{id}$ and checks if it has any write privilege or not. Then IS invokes a check to find if that $F_{id}$ exists or not. If it exists then the update method takes over, otherwise, it is sent to the fragmentation and tag generation process to partition the data into data blocks $B_i$ and create corresponding hashes $H_k$. Next, based on the length of $B_i$ and the number of maximum data servers $nS_{m}$ allocated for the user, the number of optimum servers $nS_{op}$ is calculated. $nS_{op}$, $B_i$, $H_k$, and $R_f$ are used by the subset creation algorithm \ref{algo_3} to make the optimal subsets $B_{ss}$ of blocks that could maximize the fault tolerance. Here subsets fulfill three conditions,  1) data block placement ensures maximum distance such that two copies of the same block never occur in one subset; 2) the number of subsets is strictly equal to the number of optimal servers; 3) In all subsets, each block occurs exactly $R_f$ times. Then the rating calculation is invoked (algorithm \ref{algo_rate}) to find out which subset should be sent to which data server. Rating calculation takes several factors into account: duplication count, available server storage, server load, etc. using eq: (1), and creates an overall weighted sum of rating. Then for each $B_i$,  one $S_i$ is selected such that all of the $B_i$s have different $S_i$s and the overall combination maximizes the total rating. Now, we need to send each $B_i$ to the $S_i$ by checking if $H_k$ is already in the \texttt{HashMap} ($HM$). Then $IS$ finds out the empty storage address $A_j$ for that $S_i$ and requests $S_i$ with $A_j$ to store $B_i$. $IS$ stores ($S_i,A_j$) pair with key $H_k$ in $HM$, updates the data server map by making $A_j$ address unavailable, and also stores $H_k$ in the user map to facilitate the recreation of the file in the future. 

\textbf{Update.} Users who want to update their files $F_{up}$, must first initiate four fundamental data processing operations (i.e., key generation, encryption, block creation, and tag generation). This will return data blocks and their corresponding hash values for the new files that need to be updated. After that, the $IS$ will find the updates that need to be done by users. $IS$ will achieve this by comparing old and new block hashes and saving the differences in an array called {$dif_{h}$}. Following that, the \texttt{write} function is invoked, and data is stored depending on the redundancy factor and optimal server rating calculation. First, $IS$ checks if the new blocks are duplicates or not. If the block is duplicate, it will return a block pointer. Otherwise, if the hash of the block is new, then $IS$ checks the storage availability.  Next, the data server's availability status will be adjusted so that data subsets are distributed to all servers using \texttt{write} operation and overwrite is avoided. $IS$ will then need to update the new subset location in the server and the associated server ID in the hash table. Finally, the user table will be updated to reflect the ownership of new block hashes. 

\begin{figure*} [ht]
    \centering
     
         \includegraphics[width=6.5in, height=3.5in]{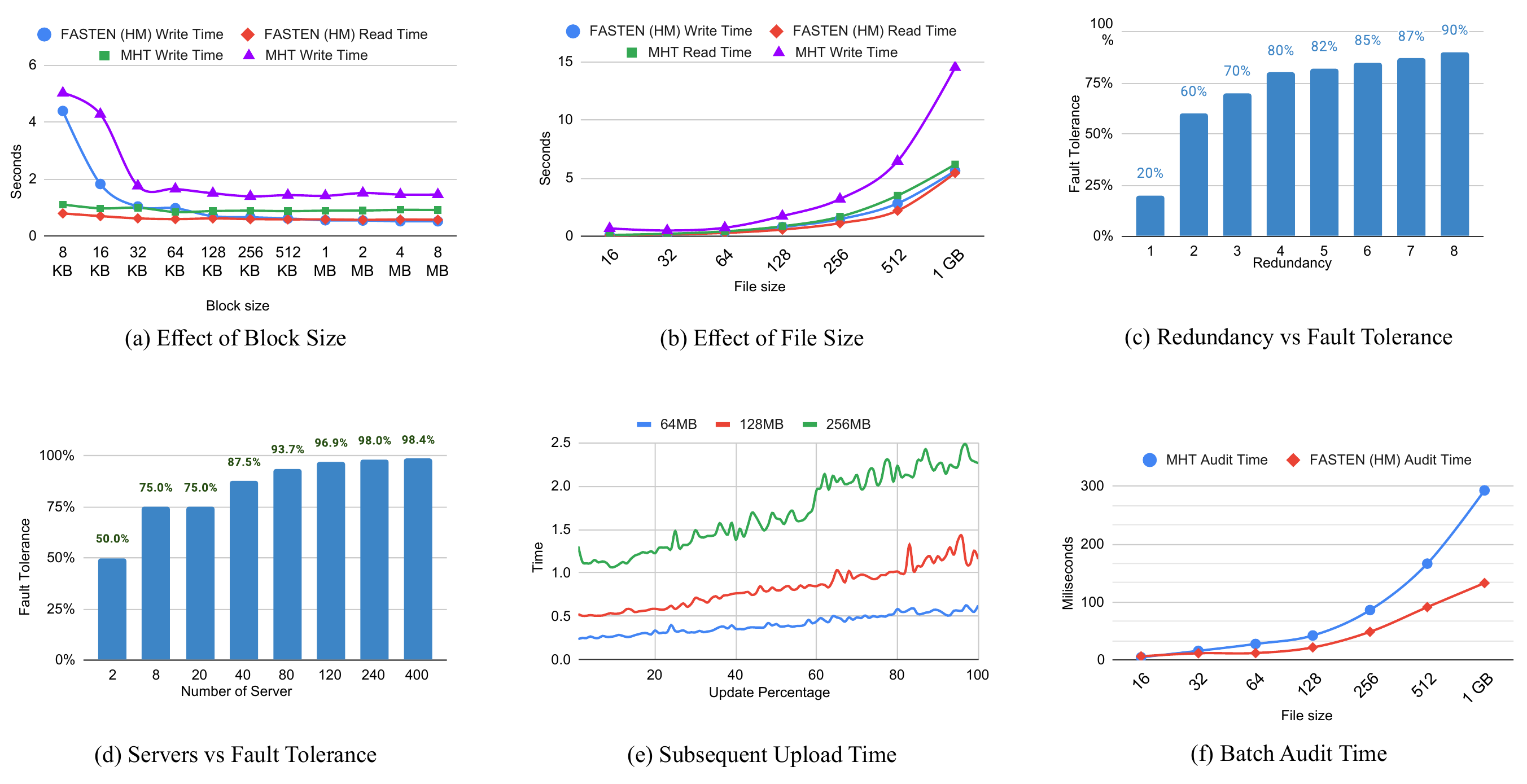}
         \caption{Performance Analysis of \texttt{FASTEN}}
         \label{fig:block} 
\end{figure*}

\textbf{Data Restoration.} For a read request, the user requests the IS with a user id ($U_{id}$) and file id ($F_{id}$). Each $IS$ validates the request, and using $U_{id}$ and $F_{id}$ finds out the sequential key from the user map data structure. Then for each key, $IS$ finds the storage addresses and requests the server to send the data blocks. This process continues for all keys in $F_{id}$ and creates the whole data from data blocks. The concatenated data is then sent to the user, where the user can decrypt and access the original data.

\textbf{Block-level Deduplication.} The index server keeps track of all the tags or hashes of blocks that have been uploaded. First, Algorithm 1 calculates the necessary blocks' hash subsets, then Algorithm \ref{algo_rate} checks each of the tags for the $\{B_i\}$s to determine the degree of duplication. It does this by utilizing a $HM_i$, which can check for the presence of any given tag in O(1) time and return the $A_j$ and $S_{id}$ of that tag. The $\{B_i\}$ must be saved if these return values are empty so that data blocks can be sent to the appropriate servers. 

\textbf{File-level Deduplication.} When an $IS$ receives a file upload request, it first checks to see if the server has the corresponding file authentication tag. If so, the server considers it a duplicate file upload request and prompts the user to use the user ID to verify file ownership. If the verification fails, the server terminates the upload activity of the file. As the block size of our system varies and is user-defined, if the user sets the block size the same as the file size, our system can perform file-level deduplication also in comparatively less time. 
\section{Implementation and Evaluation} \label{experimental} We have built a prototype of \texttt{FASTEN} in Google Colab \cite{bisong_google_2019} leveraging SHA-256 as our hash function and AES-256 for encryption. We have considered four variables to measure our performances namely, block size, redundancy factor, file size, and maximum number of allocated servers. For each test run, we kept any three variables unchanged and observed the effect of changing the other variable.

     

\subsection{Block Size vs Read-Write} In Fig. 3(a), we have presented the read-write time for varying block sizes where the file size was 128 MB, the redundancy factor was set to 3, and data was dispersed to a maximum of 40 servers. We compared the read-write performance of our Hashmap (HM) based scheme with the Merkle Hash Tree (MHT) based scheme. It is evident that, for both schemes, write time decreases as block size increases while read time remains fairly consistent. However, our HM scheme outperforms the MHT scheme for both read and write operations. 
\subsection{File Size vs Read-Write}
In Fig. 3(b), we have plotted the read-write time against varying file sizes. As file size increases, the computation time also increases. Our HM scheme once again outperforms the MHT scheme for both read and write operations. For a 1 GB file, our HM scheme is 20\% faster than the MHT scheme in read operations; and nearly 2.5X faster in write operations. 

\subsection{Redundancy Factor vs Fault-tolerance} In Fig. 3(c), we have observed fault tolerance for various redundancy factors ranging from 1 to 8 while maintaining the maximum server count of 20. We discovered that by simply keeping one copy of data, our system can withstand up to four server failures (i.e., 16 out of 20 servers need to be available to fully recover data), indicating that the system is 20\% fault tolerant ensuring 100\% availability. Similarly, if we keep 4 copies of data, our scheme achieves 80\% fault tolerance; and 90\% fault tolerance is achieved if we have 8 copies of data. 
\subsection{No. Of Servers vs Fault-tolerance} In Fig. 3(d), we have plotted fault tolerance against a varying number of servers while setting the redundancy factor to 5, file size to 128 MB, and block size to 32 KB.  We have checked fault tolerance by randomly shutting down a portion of the total servers. We observe that our system achieves 87.5\% fault tolerance when the maximum number of servers is set to 80; which means our system only needs 10 servers to recreate the original data in case of failure. If we have a total of 400 servers, fault tolerance can become as high as 98.4\%. 
\subsection{Data Update Percentage vs Time} Since the update operation does separate block-by-block matching, it takes more time than the first write. Fig. 3(e) represents times needed for a subsequent upload with 1\% to 100\% change of data blocks for file sizes of 64 MB, 128 MB, and 256 MB. The redundancy factor for this experiment was fixed at 5, the block size 32 KB, and the maximum number of servers at 20. The time requires increases with the percentages of data changes. Another intriguing pattern is that data write time for a given file ID with 100\% change is 2X slower compared to the initial write. 
\subsection{File Size vs Batch Auditing Time} In Fig. 3(f), we have shown the comparison of batch auditing time between our proposed HM and the MHT scheme. We have fixed the maximum number of servers to 40 and, the block size to 64 KB while keeping the redundancy factor at 3. We challenged the servers to check the integrity of randomly chosen 5\% data blocks for batch auditing and observed that our HM scheme performs better than the MHT scheme. For a 1 GB file, HM is at least 2X faster in auditing than the MHT.

\section{Comparison With Related Work} \label{related}

\textbf{Deduplication.} Sing et al. \cite{singh2018secure} proposed a convergent encryption-based deduplication scheme to address a single point of failure, but it involves computationally heavy tasks due to splitting data into numerous shares. This solution will therefore fail in the event of a cloud service provider lockouts. Yuan et al. \cite{yuan2019secure} suggested a secure deduplication scheme using re-encryption and a bloom filter-based location selection mechanism, but multiple stages of encryption make it complex. In contrast, our proposed architecture ensures lightweight deduplication at both file and block levels.

\begin{figure}[ht]
\centering
\includegraphics[width=3in, height=1.8in]{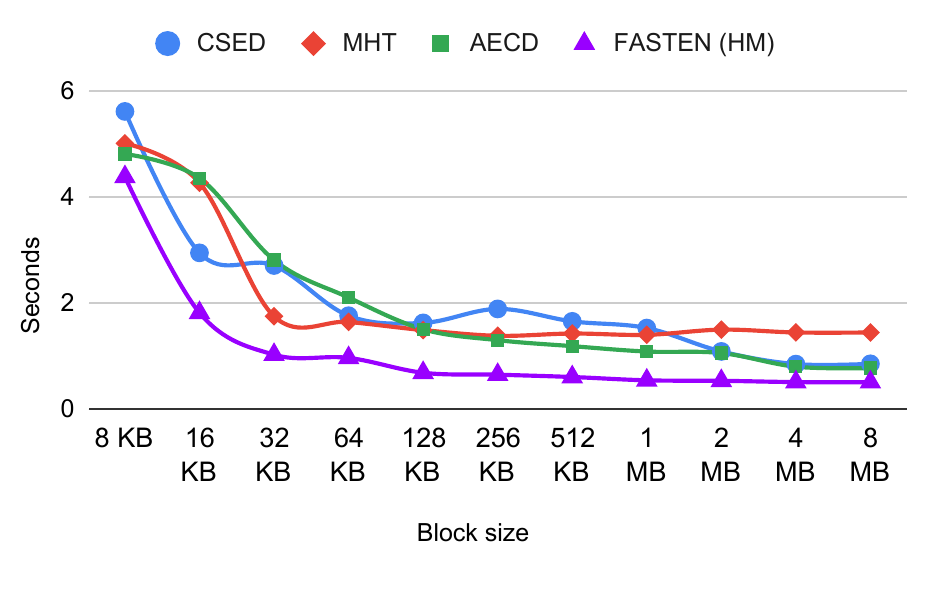} 
\caption{Deduplication write time Comparison}
\label{dcomp}
\end{figure}

\begin{table}[ht]
\renewcommand{\arraystretch}{1.2}
\caption{Deduplication Performance Comparison}
\label{tab:dedup}
\centering
\setlength{\tabcolsep}{4pt}
\begin{tabular} {|c||c|c|c||c|}
\hline 
\bf  Deduplicated Blocks  & \bf CSED & \bf MHT  & \bf AECD & \bf FASTEN (HM) \\
\hline
4096 & 85.79\% & 94.82\% & 83.59\% & 87.93\%   \\
\hline
2048 & 87.25\% & 88.86\% & 79.44\% & 88.37\%    \\
\hline
1024 & 86.03\% & 92.57\% & 86.52\% & 86.03\%     \\
\hline  
\end{tabular}
\end{table}


Li et al.\cite{li2019csed} proposed CSED, a client-side deduplication scheme for centralized servers, but it lacks feasibility in large-scale cloud platforms with multiple redundant servers. Moreover, it is vulnerable to adversaries impersonating valid clients. Yang et al.\cite{yang2020achieving} introduced AECD, an efficient access control secure deduplication scheme using Boneh-Goh-Nissim cryptography and a bloom filter data structure. However, additional computation increases the time complexity of the AECD system~\cite{yang2020achieving}.

\begin{figure}[ht]
\centering
\includegraphics[width=2.4in, height=1.3in]{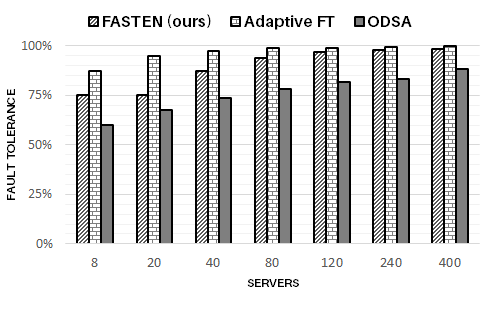} 
\caption{Fault Tolerance Performance Comparison} 
\label{fcomp}
\end{figure}

In Figure \ref{dcomp}, we compared our deduplication efficiency with CSED, AECD, and MHT-based algorithms, maintaining a constant environment. By varying block sizes, we demonstrated our scheme's superior speed. Unlike other schemes focused on storage efficiency, our approach ensures both storage cost savings and higher availability, addressing redundancy control for comprehensive deduplication benefits.

\begin{figure}[ht]
\centering
\includegraphics[width=2.9in, height=1.6in]{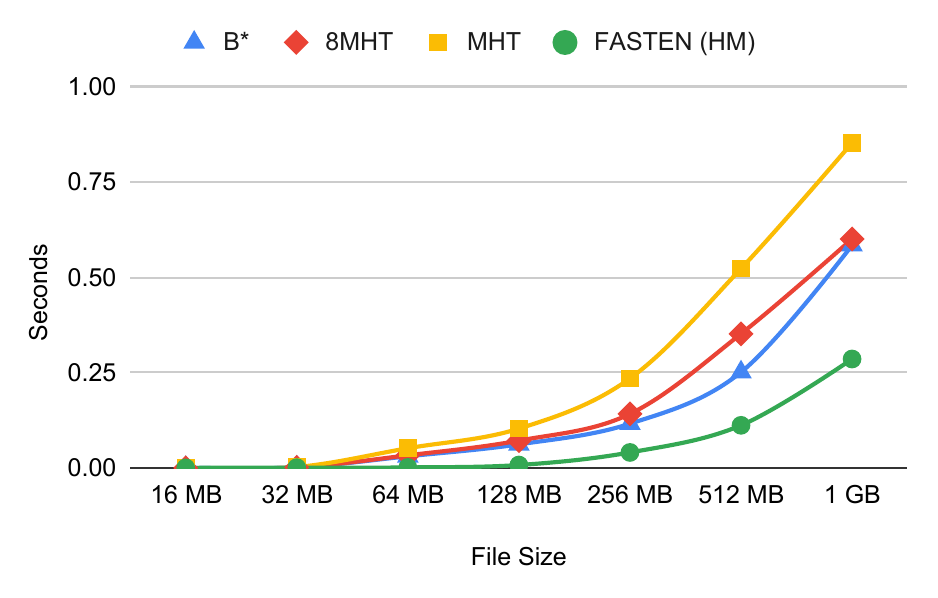} 
\caption{Batch Audit Performance Comparison} 
\label{acomp}
\end{figure}


Table \ref{tab:dedup} shows the percentage of data blocks that have been successfully removed from the data server. The MHT-based approach mostly uses one tree to store the values thus its percentage of duplicated blocks is higher than other approaches. However, a large number of blocks can saturate the data server with a large set of trees, hence increasing the overall seek time. CSED and AECD both used one centralized server system thus limiting their computation ability to perform multiple searches. Compared to these two methods our proposed deduplication works 1\% to 2\% better in removing duplicated blocks. 


\textbf{Fault Tolerance.} Sathiyamoorthi et al. \cite{sathiyamoorthi2021adaptive} proposed adaptive resource allocation with a fault tolerance detector, outperforming \texttt{FASTEN} for fault tolerance when servers are fewer than 40 in a predicted environment. However, the approach's impracticality lies in predicting and caching, oversimplifying the problem. Both \texttt{FASTEN} and Adaptive FT perform similarly, achieving above 98\% when there are more than 120 servers. Notably, Adaptive FT focuses on availability and fault tolerance but overlooks data deduplication. On the other hand, Wang et al. \cite{wang2020optimized} proposed a fault tolerance strategy using a Gaussian random generator but faced limitations in predicting data access and storage configuration times. This resulted in lower fault tolerance compared to \texttt{FASTEN} as server numbers increased. Additionally, the scheme lacked clarity on how the ratio of access time and storage configuration contributes to fault tolerance optimization.

\textbf{Batch Audit.} Since \texttt{FASTEN} incorporates batch auditing using an HM data structure, it is compared with similar methods in Fig. \ref{acomp}. Luo et al.'s B* system \cite{luo2021mhb}, integrating B tree with MHT, achieves lower time complexity than MHT and 8 MHT \cite{yue2020blockchain} by concentrating all data in leaf nodes. In contrast, MHT and 8 MHT exhibit more complex time complexities ($O(\log_{2}(N))$ and $O(\log_{8}(N))$, respectively), whereas our proposed \texttt{FASTEN} scheme maintains a time complexity of $O(constant)$.


All the existing schemes prioritize either availability or storage optimization, but our proposed scheme strikes a balance, offering user-defined availability and fault tolerance with an optimal number of redundant servers.


\section{Conclusion} \label{conclusion} 
In this paper, we propose \texttt{FASTEN}, an efficient deduplication scheme for cloud storage in redundant servers scenarios. \texttt{FASTEN} ensures storage efficiency, high availability, and reliability using a custom \texttt{HashMap} with convergent encryption. It allows users to set the redundancy factor by selecting the best available servers. Our custom data structure, employing a pair-matching algorithm, produces server ratings. An index server mediates redundancy and deduplication balance. We implemented a \texttt{FASTEN} prototype, measuring \texttt{read}, \texttt{write}, \texttt{update}, \texttt{batch auditing}, and \texttt{fault-tolerance} performance for varying file and block sizes. Comparison with an MHT-based scheme demonstrates \texttt{FASTEN}'s efficiency and reduced computation time. We also compare deduplication, fault tolerance, and batch auditing performance with existing schemes, showing superior or comparable results.

\bibliographystyle{IEEEtran}
\bibliography{IEEEabrv,mybibfile}
\end{document}